\documentclass[aps,prb,reprint,superscriptaddress,citeautoscript,floatfix]{revtex4-1}
\usepackage{amsmath}
\usepackage{graphicx}
\usepackage{epstopdf}
\usepackage{hyperref}
\usepackage{color,soul}

\usepackage{array}
\usepackage{booktabs}
\usepackage{multirow}

\setstcolor{red}
\usepackage[normalem]{ulem}
\usepackage{xcolor}
\newcommand\rst{\bgroup\markoverwith{\textcolor{red}{\rule[0.5ex]{2pt}{1.5pt}}}\ULon}

\let\oldhat\hat

\renewcommand{\hat}[1]{\oldhat{\mathbf{#1}}}
\newcommand{\upperRomannumeral}[1]{\uppercase\expandafter{\romannumeral#1}}

\makeatletter
\def\hlinewd#1{%
  \noalign{\ifnum0=`}\fi\hrule \@height #1 \futurelet
   \reserved@a\@xhline}
\makeatother

\begin{document}


\title{Microwave Meissner Screening Properties of Proximity coupled Topological Insulator / Superconductor Bilayers }


\author{Seokjin Bae}
\affiliation{Center for Nanophysics and Advanced Materials, Department of Physics, University of Maryland, College Park, MD 20742, USA}

\author{Seunghun Lee}
\affiliation{Center for Nanophysics and Advanced Materials, Department of Physics, University of Maryland, College Park, MD 20742, USA}
\affiliation{Department of Materials Science and Engineering, University of Maryland,
College Park, Maryland 20742, USA}

\author{Xiaohang Zhang}
\affiliation{Center for Nanophysics and Advanced Materials, Department of Physics, University of Maryland, College Park, MD 20742, USA}
\affiliation{Department of Materials Science and Engineering, University of Maryland,
College Park, Maryland 20742, USA}

\author{Ichiro Takeuchi}
\affiliation{Center for Nanophysics and Advanced Materials, Department of Physics, University of Maryland, College Park, MD 20742, USA}
\affiliation{Department of Materials Science and Engineering, University of Maryland,
College Park, Maryland 20742, USA}

\author{Steven M. Anlage}
\affiliation{Center for Nanophysics and Advanced Materials, Department of Physics, University of Maryland, College Park, MD 20742, USA}

\date{\today}

\begin{abstract}
The proximity coupled topological insulator / superconductor (TI/SC) bilayer system is a representative system to realize topological superconductivity. In order to better understand this unique state and design devices from the TI/SC bilayer, a comprehensive understanding of the microscopic properties of the bilayer is required. In this work, a microwave Meissner screening study, which exploits a high-precision microwave resonator technique, is conducted on the SmB$_6$/YB$_6$ thin film bilayers as an example TI/SC system. The study reveals spatially dependent electrodynamic screening response of the TI/SC system that is not accessible to other techniques, from which the corresponding microscopic properties of a TI/SC bilayer can be obtained. The TI thickness dependence of the effective penetration depth suggests the existence of a bulk insulating region in the TI layer. The spatially dependent electrodynamic screening model analysis provides an estimate for the characteristic lengths of the TI/SC bilayer: normal penetration depth, normal coherence length, and the thickness of the surface states. We also discuss implications of these characteristic lengths on the design of a vortex Majorana device such as the radius of the vortex core, the energy splitting due to intervortex tunneling, and the minimum thickness required for a device. 
\end{abstract}


\maketitle

\section{Introduction}
\par Creating an experimental platform which hosts Majorana bound states (MBSs) in a condensed matter system is a goal that has received great attention recently.\cite{Alicea2012RPP,Beenakker2013ARCMP} Due to robust topological protection, the MBS is a promising qubit candidate for quantum computation.\cite{Kitaev2006AP} One of the platforms proposed to realize the MBS is a topological insulator / superconductor (TI/SC) bilayer system.\cite{LiangFu2008PRL} With the induced chiral $p$-wave superconductivity in the topological surface states (TSS), an MBS has been predicted to exist in its vortex core.\cite{Read2000PRB,Ivanov2001PRL,Stern2004PRB,Stone2006PRB} Therefore, it is important for the physics community to establish and understand the properties of TI/SC bilayer systems.

\par There have been a number of studies on the Bi-based TI (Bi$_2$Se$_3$, Bi$_2$Te$_3$, etc) /SC systems through point contact spectroscopy (PCS)\cite{WenqingDai2017SciRep}, ARPES\cite{MWang2012Science,SuYangXu2014NatPhys}, and STM\cite{JinPengXu2014PRL,JinPengXu2015PRL,HaoHuaSun2016PRL} measurements. PCS and STM probe the magnitude of the superconducting order parameter induced in the top surface of the TI with a probing depth range limited to the mean free path or coherence length, and cannot be applied to the case when an insulating bulk region is present. ARPES studies the angle-resolved magnitude of the induced order parameter from the first few atomic layers of the top surface of the TI. 

\par In contrast, a microwave Meissner screening study investigates the high frequency electromagnetic field response. The microwave field propagates through an insulating layer and penetrates inside the superconducting system to the scale of the penetration depth, which is comparable to the thickness of typical thin-film bilayers ($< 200$ nm). Since the field screening response arises throughout the entire bilayer, it can reveal more details of the proximity-coupled bilayer\cite{Deutscher1969,Pambianchi1994PRB,Belzig1996PRB,deGennes1999,JKim2005PRB} that are not directly available to the other techniques. It is also important to note that the screening response study does not require specialized surface preparation which is critical for many of the other techniques.

\par The distinct capabilities of the Meissner screening study on the proximity-coupled system have been previously demonstrated on conventional normal (N) / superconductor (S) bilayer systems such as Cu (N) / Nb (S).\cite{Hook1976JLTP,Simon1984PRB,Kanoda1987PRB,Mota1989JLowTemp,Claassen1991PRB,Pambianchi1994PRB,Pambianchi1995PRB,Onoe1995JPSJ,Pambianchi1996PRB,Pambianchi1996PRB2} It can reveal the spatial distribution of the order parameter and the magnetic field profile throughout the film, as well as their evolution with temperature. From such information, superconducting characteristic lengths such as the normal coherence length $\xi_\text{N}$ and normal penetration depth $\lambda_\text{N}$ of the proximity-coupled normal layer can be estimated. The study can also reveal thickness dependent proximity-coupling behavior, which helps to estimate the thickness of the surface states ($t_{\text{TSS}}$) for TI/SC bilayers. The $\xi_\text{N}$, $\lambda_\text{N}$, and $t_\text{TSS}$ of a proximity-coupled TI layer determine the radius of a vortex, the maximum spacing between vorticies in a lattice, and the minimum thickness of the TI layer. Such information is required to avoid intervortex tunneling of MBSs, which would result in a trivial fermionic state.\cite{MCheng2010PRB}

\par Compared to other high frequency electromagnetic techniques such as THz optical measurement, the advantage of the microwave Meissner screening study for investigating the properties of a TI/SC bilayer is that the energy of a 1 GHz microwave photon ($\approx 4$ $\mu$eV) is a marginal perturbation to the system. On the other hand, the energy of a 1 THz optical photon ($\approx 4$ meV) is comparable to the gap energy ($\leq 3$ meV) of typical superconductors used in TI/SC systems such as Nb, Pb, Al, NbSe$_2$, and YB$_6$.\cite{Kittel,Clayman1971SSC,Kadono2007PRB} Therefore, the microwave screening study is an ideal method to study details of the induced order parameter in TI/SC bilayers.

\par In this article, we conduct a microwave Meissner screening study on SmB$_6$/YB$_6$: a strong candidate for topological Kondo insulator / superconductor bilayer systems. The existence of the insulating bulk in SmB$_6$ is currently under debate.\cite{Menth1969PRL,NXu2014NatComm,Syers2015PRL,Tan2015Science,Laurita2016PRB,YXu2016PRL,JingdiZhang2018PRB,YunsukEo2018ArXiv} From measurements of the temperature dependence of the Meissner screening with a systematic variation of SmB$_6$ thickness, this study shows evidence for the presence of an insulating bulk region in the SmB$_6$ thin films.  Through a model of the electrodynamics, the study also provides an estimation for the characteristic lengths of the bilayer system including the thickness of the surface states.

\section{Experiment}
\par
\begin{figure}
		\includegraphics[width=1\columnwidth]{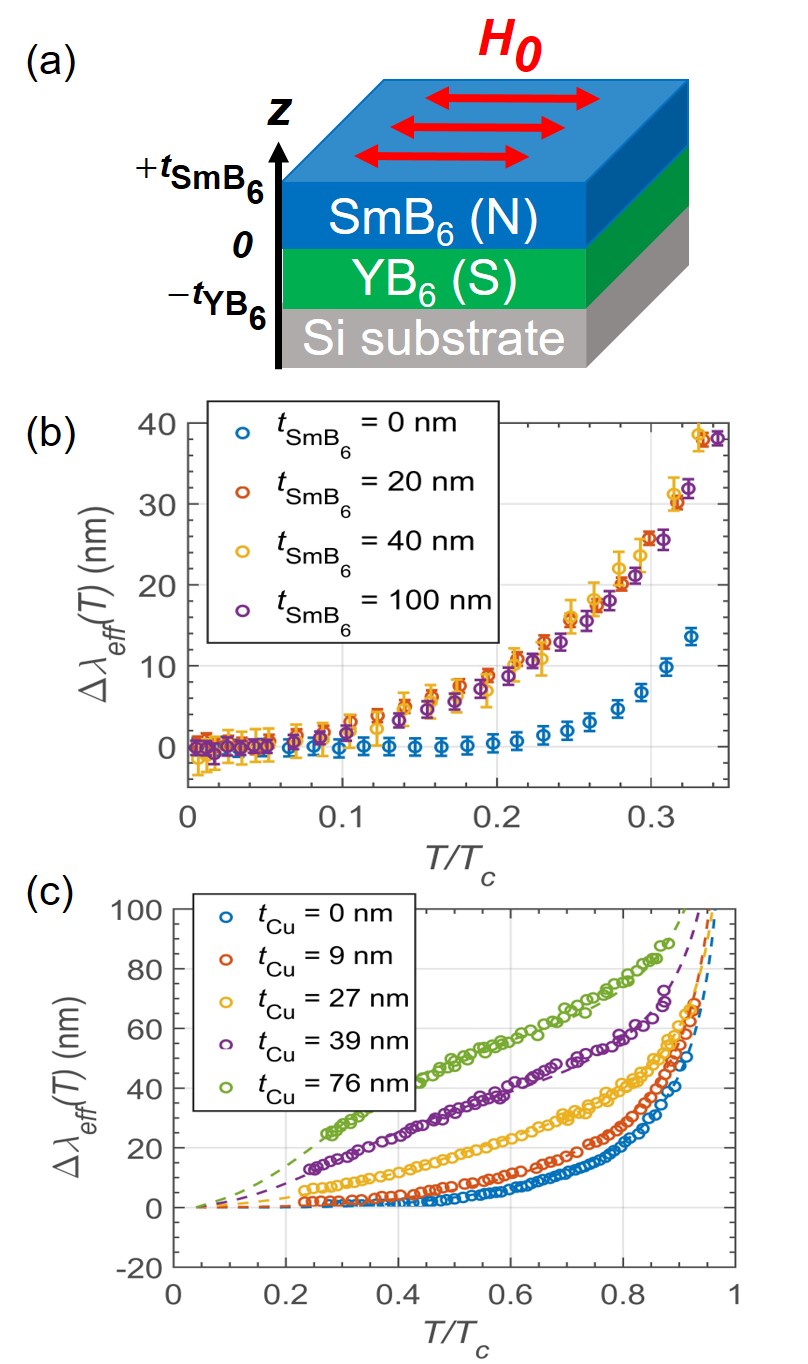}
		\caption{\label{fig:Fig1} (a) A schematic of the bilayer consisting of an SmB$_6$ film and a YB$_6$ film. A parallel microwave magnetic field ($H_0$) is applied to the top surface of the SmB$_6$ layer (red arrows). (b) Temperature dependence of the effective penetration depth $\Delta\lambda_{eff}(T)$ of the SmB$_6$/YB$_6$ bilayers for various SmB$_6$ layer thickness ($t_{\text{SmB}_6}$). (c) $\Delta\lambda_{eff}(T)$ of a Cu/Nb (conventional metal / superconductor) bilayers\cite{Pambianchi1996PRB} for various Cu layer thickness ($t_{\text{Cu}}$). The dashed lines are the model fits.\cite{Pambianchi1996PRB} }
\end{figure}

 SmB$_6$/YB$_6$ bilayers were fabricated through a sequential sputtering process without breaking the vacuum to ensure a pristine interface between SmB$_6$ and YB$_6$ for ideal proximity coupling. The details of sample fabrication can be found in the Appendix Sec. \ref{SampleGrowth}. The geometry of the bilayers is schematically shown in Fig. \ref{fig:Fig1}(a). The YB$_6$ film has a thickness of 100 nm and $T_c=6.1$ K obtained from a DC resistance measurement.\cite{SeunghunLee2019Nature} The thickness of SmB$_6$ layers ($t_{\text{SmB}_6}$) are varied from 20 to 100 nm for systematic study. These bilayers all have $T_c=5.8\pm0.1$ K without a noticeable $t_{\text{SmB}_6}$ dependence of $T_c$. The measurement of the effective penetration depth $\lambda_{eff}$ is conducted with a dielectric resonator setup.\cite{HakkiColeman1960,Mazierska1998IEEE,SeokjinBae2019RSI} A 3 mm diameter, 2 mm thick rutile (TiO$_2$) disk, which facilitates a microwave transmission resonance at 11 GHz, is placed on top of the sample mounted in a Hakki-Coleman type resonator.\cite{HakkiColeman1960} This resonator consists of niobium (top) and copper (bottom) plates to obtain a high quality factor for the dielectric resonance. The resonator is cooled down to the base temperature of 40 mK. As the temperature of the sample is increased from the base temperature, the change of the resonance frequency is measured, $\Delta f_0(T)=f_0(T)-f_0(T_{ref})$. $T_{ref}$ here is set to $230$ mK ($\approx 0.04T_c$ of the bilayers), below which $f_0(T)$ of the bilayers shows saturated temperature dependence. This data is converted to the change in the effective penetration depth $\Delta\lambda_{eff}(T)$ using a standard cavity perturbation theory,\cite{Klein1992JSuper,BBJin2002PRB,Ormeno2002PRL}
\begin{equation}
\Delta\lambda_{eff}(T) =\lambda_{eff}(T)-\lambda_{eff}(T_{ref})= -\frac{G_{geo}}{\pi \mu_0}\frac{\Delta f_0(T)}{f_0^2(T)}.
\end{equation}
Here, $G_{geo}$ is the geometric factor of the resonator.\cite{SeokjinBae2019RSI}

\par Fig. \ref{fig:Fig1}(b) shows $\Delta\lambda_{eff}(T)$ for the SmB$_6$ (N) / YB$_6$ (S) bilayers for various SmB$_6$ layer thickness $t_{\text{SmB}_6}$. The single layer YB$_6$ thin film (i.e., $t_{\text{SmB}_6}=0$) shows temperature independent behavior below $T/T_c<0.2$. This is  not only consistent with the BCS temperature dependence of $\Delta\lambda(T)$ for a spatially homogeneous, fully-gapped superconductor,\cite{Abrikosov1988,Prozorov2006SST} but also consistent with previous observations on YB$_6$ single crystals.\cite{Kadono2007PRB,Tsindlekht2008PRB} However, once the SmB$_6$ layer is added, $\Delta\lambda_{eff}(T)$ clearly shows temperature dependence below $T/T_c<0.2$. Here, the important unconventional feature is that the low temperature profile of $\Delta\lambda_{eff}(T)$ for the SmB$_6$/YB$_6$ bilayers shows only a marginal $t_{\text{SmB}_6}$ dependence. This is in clear contrast to the case of the Cu (N) / Nb (S) bilayers shown in Fig. \ref{fig:Fig1}(c). The $\Delta\lambda_{eff}(T)$ for this conventional metal/superconductor bilayer system shows considerable evolution as the normal layer thickness $t_{\text{Cu}}$ increases. This is because when the decay length of the induced order parameter $\xi_\text{N}(T)$ decreases with increasing temperature, the thicker (thinner) normal layer undergoes a larger (smaller) change in the spatial distribution of the order parameter, and hence the spatial profile of the screening. Therefore, the marginal $t_{\text{SmB}_6}$ dependence of $\Delta\lambda_{eff}(T)$ for the SmB$_6$/YB$_6$ bilayer implies that even though $t_{\text{SmB}_6}$ is increased, the actual thickness of the proximity-coupled screening region in the SmB$_6$ layer remains roughly constant.

\section{Model}
\par To quantitatively analyze this unconventional behavior, an electromagnetic screening model for a proximity-coupled bilayer is introduced.\cite{Pambianchi1994PRB,Pambianchi1995PRB,Pambianchi1996PRB,Pambianchi1996PRB2} The model solves Maxwell's equations combined with the second London equation for the current and field inside the bilayer with appropriate boundary conditions at each temperature (See Appendix \ref{Screening model equation}), to obtain the spatial profile of the magnetic field $H(z,T)$ and the current density $J(z,T)$ as a function of temperature,\cite{Pambianchi1994PRB} where $z$ denotes the coordinate along the sample thickness direction as depicted in Fig. \ref{fig:Fig1}(a). From the obtained field and current profiles, one can obtain the total inductance $L(T)$ of the bilayer as
\begin{equation} \label{inductance}
\begin{split}
L(T) &=\frac{\mu_0}{H_0^2}  \int_{-t_\text{S}}^{0}  \left[H^2(z,T)+\lambda_\text{S}^2(T)J^2(z,T)\right]dz \\
&+ \frac{\mu_0}{H_0^2}\int_{0}^{+d_\text{N}} \left[H^2(z,T)+\lambda_\text{N}^2(z,T)J^2(z,T)\right]  dz \\
&+ \frac{\mu_0}{H_0^2}\int_{+d_\text{N}}^{+t_\text{N}} \left[H^2(z)\right]  dz,
\end{split}
\end{equation}
from which one can obtain an effective penetration depth from the relation $L(T)=\mu_0\lambda_{eff}(T)$. Here, $H_0$ is the amplitude of the applied microwave magnetic field at the top surface of the normal layer (see Fig.  \ref{fig:Fig1}(a)), $\lambda_\text{S}$ ($\lambda_\text{N}$) is local penetration depth of the superconductor (normal layer), $t_\text{S}$ is the thickness of the superconductor, $t_\text{N}$ (N=SmB$_6$ or Cu) is the total thickness of the normal layer, and $d_\text{N}$ ($\leq t_\text{N}$, integration limit of the second and third terms in Eq. (\ref{inductance})) is the thickness of the proximity-coupled region in the normal layer, which is assumed to be temperature independent. In Eq.(\ref{inductance}), $H^2$ is proportional to field stored energy and $\lambda^2J^2$ is proportional to kinetic stored energy of the supercurrent. The first, second, and third integration terms come from the superconductor, the proximity-coupled part of the normal layer, and the uncoupled part of the normal layer, respectively. 

\par
\begin{figure}
		\includegraphics[width=1\columnwidth]{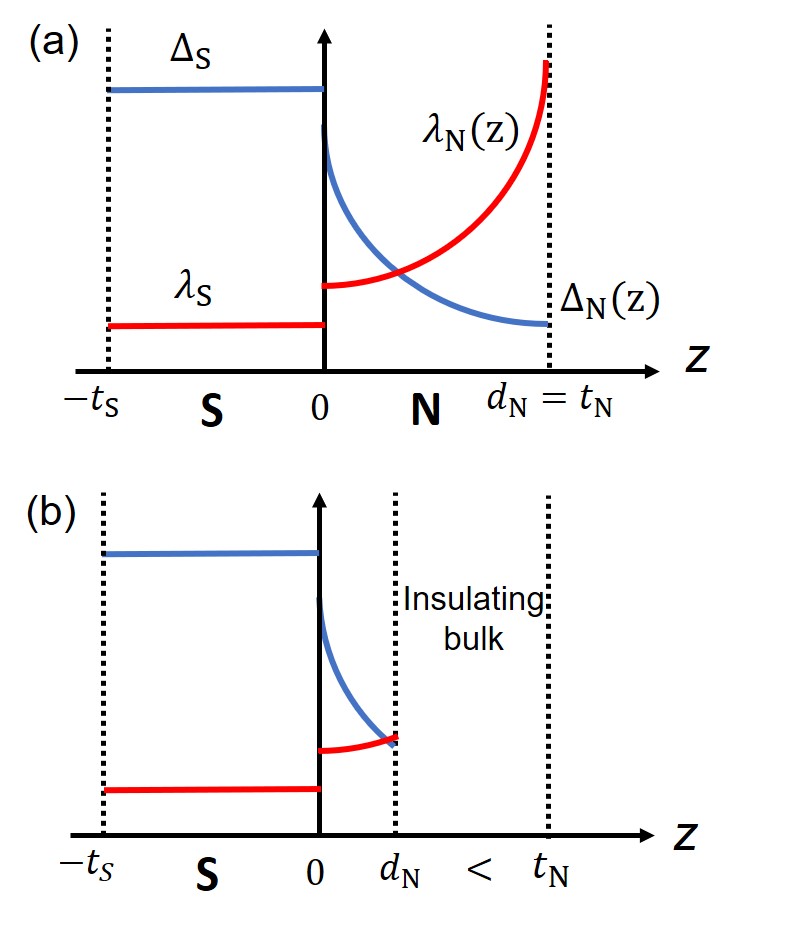}
		\caption{\label{fig:Fig2} (a) Schematic spatial profile of the order parameter $\Delta_\text{N,S}$ (blue) and the local penetration depth $\lambda_\text{N,S}$ (red) through the normal layer (N) / superconductor (S) bilayer sample for the case of the absence of an insulating bulk. $z$ is the thickness direction coordinate and $t_\text{N}$ ($t_\text{S}$) is the thickness of the normal layer (superconductor). The proximitized thickness $d_\text{N}$ is equal to the normal layer thickness $t_\text{N}$. (b) In the presence of an insulating bulk, $d_\text{N} < t_\text{N}$ since the insulating bulk blocks propagation of the order parameter to the top surface. Note that the microwave magnetic field is applied to the right surfaces. }
\end{figure}

 A schematic view of the order parameter profile in the bilayers is shown in Fig. \ref{fig:Fig2}. As seen in Fig. \ref{fig:Fig2}(a), for a conventional metal, $d_\text{N}$ is the same as $t_\text{N}$ since the entire normal layer is uniformly susceptible to induced superconductivity, and thus the third integration term in Eq. \ref{inductance} becomes zero. However, as seen in Fig. \ref{fig:Fig2}(b), if there exists an insulating bulk region blocking the propagation of the order parameter up to the top surface in the normal layer (as in the case of a thick TI), only the bottom conducting surface adjacent to the superconductor is proximity-coupled. In this case, $d_\text{N}$ becomes the thickness of the bottom conducting surface states. The third integration term in Eq. (\ref{inductance}), which accounts for the uncoupled portion of the normal layer, becomes non-zero. However, this third term can be removed by taking $\Delta L(T)$ into account since the un-coupled SmB$_6$ region has temperature-independent microwave properties below 3 K\cite{Sluchanko2000PRB}, whereas the temperature range of the measurement here extends below 2 K.

\par The spatial dependence of screening of the proximity-coupled normal layer is imposed by that of the induced order parameter $\Delta_\text{N}$ (Fig. \ref{fig:Fig2}(a)), which can be approximated by an exponential decay profile $\Delta_\text{N}(z,T)=\Delta_\text{N}(0,T)e^{-z/\xi_\text{N}(T)}$ in terms of the normal coherence length $\xi_\text{N}(T)$.\cite{deGennes1999} The position dependent normal penetration depth is inversely proportional to the order parameter $\lambda_\text{N}\sim1/\Delta_\text{N}$\cite{Deutscher1969JChemSol} so its position dependence is expressed as $\lambda_\text{N}(z,T)=\lambda_\text{N}(0,T)e^{z/\xi_\text{N}(T)}$. Here, the temperature dependence of $\lambda_\text{N}$ at the interface is assumed to follow that of the superconductor\cite{Simon1981PRB} $\lambda_\text{N}(0,T)/\lambda_\text{N}(0,0)=\lambda_\text{S}(T)/\lambda_\text{S}(0) \cong 1+\sqrt{\pi\Delta_0/2k_BT}\exp(-\Delta_0/k_BT)$, which is the asymptotic behavior below 0.3$T_c$ for a fully-gapped superconductor.\cite{Abrikosov1988,Prozorov2006SST} 

\par For the temperature dependence of the screening in the normal layer, $\xi_\text{N}(T)$ plays a crucial role since it determines the spatial distribution of $\Delta_\text{N}(z,T)$. If the sample is in the clean limit, the temperature dependence of the normal coherence length is given by $\xi_\text{N} = \hbar v_F/2\pi k_B T$, where $v_F$ denotes the Fermi velocity of the N layer. In the dirty limit, it is given by $\xi_\text{N} = \sqrt{\hbar v_F l_\text{N}/6\pi k_B T}$,\cite{Deutscher1969} where $l_\text{N}$ denotes the mean-free path of the N layer. For the model fitting, the simplified expressions $\xi^{clean}_\text{N}(T)=\xi^{clean}_\text{N}(T_0)\times T_0/T$ and $\xi^{dirty}_\text{N}(T)=\xi^{dirty}_\text{N}(T_0)\times \sqrt{T_0/T}$ are used, with $\xi_\text{N}(T_0)$ as a fitting parameter. Here, $T_0$ is an arbitrary reference temperature of interest. Note that the divergence of $\xi_\text{N}(T)$ as $T\rightarrow0$ should be cut off below a saturation temperature due to the finite thickness of the normal layer, which is theoretically predicted,\cite{Falk1963,Deutscher1969} and also experimentally observed from magnetization studies on other bilayer systems.\cite{Mota1989JLowTemp,Onoe1995JPSJ} In our measurements, the effect of this saturation of $\xi_\text{N}(T)$ can be seen from the sudden saturation of the $\Delta\lambda_{eff}(T)$ data below $0.04T_c$ (see Fig. \ref{fig:Fig1}(b) and Fig. \ref{fig:Fig3}(b-d)). Therefore, only the data obtained in a temperature range of $T/T_c\geq0.04$ is fitted, where the $\Delta\lambda_{eff}(T)$ data indicates that $\xi_\text{N}$ is temperature dependent.

\par A given set of these parameters $\lambda_\text{S}(0)$, $\lambda_\text{N}(0,0)$, $\xi_\text{N}(T_0)$, and $d_\text{N}$ determines a model curve of $\Delta\lambda_{eff}(T)$. Therefore, by fitting the experimental data to a model curve, one can determine the values of these characteristic lengths. This screening model has successfully described $\Delta\lambda(T)$ behavior of various kinds of normal/superconductor bilayers.\cite{Pambianchi1995PRB,Pambianchi1996PRB,Pambianchi1996PRB2} 

\section{Results} \label{Result}
\par
\begin{figure}
		\includegraphics[width=1\columnwidth]{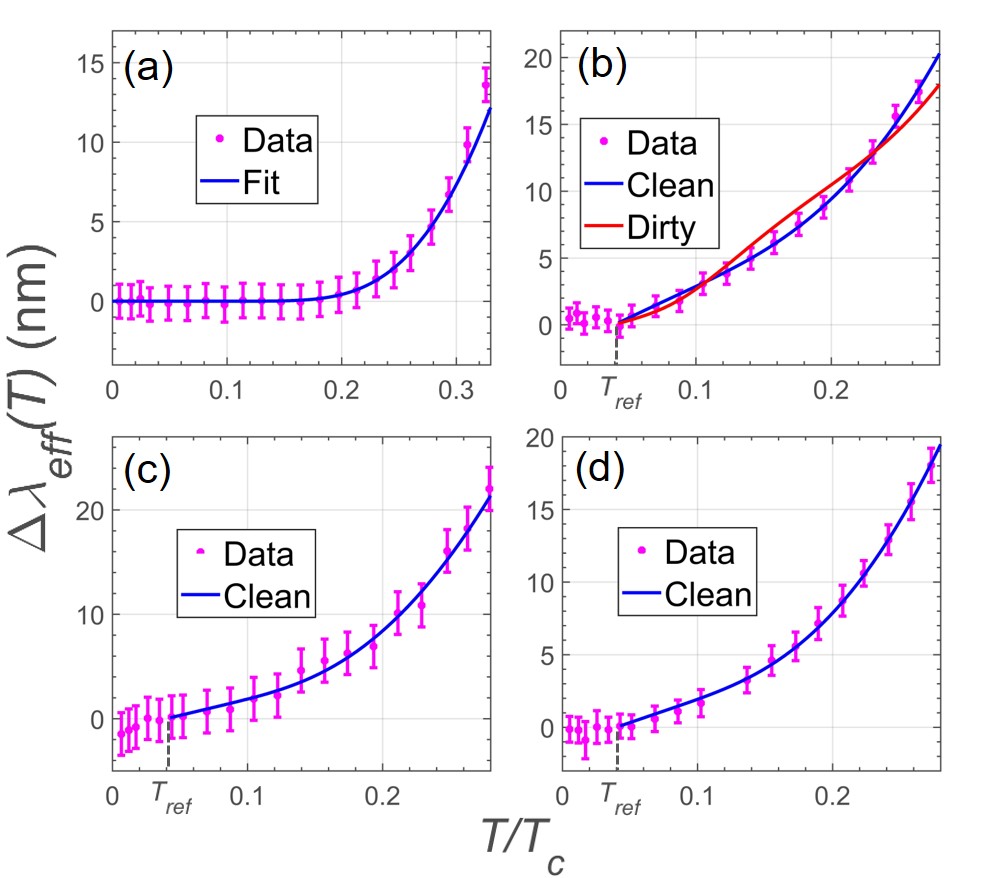}
		\caption{\label{fig:Fig3} $\Delta\lambda_{eff}(T)$ vs. $T/T_c$ data and fits for SmB$_6$/YB$_6$ bilayers at low temperature, $T/T_c<0.3$. (a) The single layer YB$_6$ (100 nm) ($t_{\text{SmB}_6}=0$ nm). The magenta points are data, and the blue line is a fit from the electromagnetic screening model. (b) The bilayer with $t_{\text{SmB}_6}=20$ nm. The blue line is a fit with the clean limit temperature dependence of $\xi_\text{N}(T)$, and the red line is a fit with the dirty limit temperature dependence. (c) and (d) The bilayers with $t_{\text{SmB}_6}=40$ nm and $100$ nm, respectively.}
\end{figure}

 As seen in Fig. \ref{fig:Fig3}(a), the model is first applied to fit $\Delta\lambda_{eff}(T)$ of a single layer YB$_6$ thin film (i.e., no SmB$_6$ layer on the top) to obtain $\lambda_\text{S}(0)$: the simplest case where one needs to consider only the first term in Eq. (\ref{inductance}). Here, the data in a temperature range of $T < 1.6$ K ($\approx 0.28 T_c$ of the SmB$_6$/YB$_6$ bilayers) is fitted to avoid the contribution from the niobium top plate to $\Delta f_0(T)$. The best fit is determined by finding the fitting parameters that minimize the root-mean-square error $\sigma$ of $\Delta\lambda_{eff}(T)$ between the experimental data and the model fit curves. The best fit gives $\lambda_\text{S}(0)=227 \pm 2$ nm (The determination of the error bar is described in Appendix \ref{ErrorbarDet}). A comparison between the estimated $\lambda_\text{S}(0)$ of the YB$_6$ thin film and that obtained in other work is discussed in the Appendix \ref{YB6 penetration depth}

\par We now fix the value of $\lambda_\text{S}(0)$ of the YB$_6$ layer and focus on extracting the characteristic lengths of the induced superconductivity of the bilayers. Recent PCS measurements on a series of SmB$_6$/YB$_6$ bilayers\cite{SeunghunLee2019Nature} help to reduce the number of fitting parameters: the point contact measurement on the bilayer with $t_{\text{SmB}_6}=20$ nm at 2 K showed perfect Andreev reflection, i.e., conductance doubling at the interface between a metal tip and the top surface of the SmB$_6$, indicating that the entire 20 nm thick SmB$_6$ layer is proximity-coupled. Therefore, $d_\text{N}$ is fixed to 20 nm when fitting the $\Delta\lambda_{eff}(T)$ data of the bilayer with $t_{\text{SmB}_6}=20$ nm.

\par The fitting is conducted with the clean and the dirty limit temperature dependence of $\xi_\text{N}(T)$ as shown in Fig. \ref{fig:Fig3}(b). The clean limit fit (blue) gives $\xi^{clean}_\text{N}(2\text{K})=52\pm1$ nm, $\lambda_\text{N}(0,0)=340\pm2$ nm with $\sigma$ of 0.237. On the other hand, the dirty limit fit (red) gives $\xi^{dirty}_\text{N}(2\text{K})=262\pm180$ nm, $\lambda_\text{N}(0,0)=505\pm7$ nm with $\sigma$ of 0.780. According to the fitting result, not only does the dirty limit fit apparently deviate from the data points, but also the $\sigma$ of the dirty limit is three times larger than that of the clean limit, implying that the clean limit is more appropriate for describing $\xi_\text{N}(T)$ of the SmB$_6$ layer. Henceforth, the $\Delta\lambda_{eff}(T)$ data for the bilayers with other $t_{\text{SmB}_6}$ is fit using the clean limit temperature dependence of $\xi_\text{N}$. Also, the obtained value of $\xi_\text{N}$(2K) = 52 nm will be used when the data of the bilayers with other $t_{\text{SmB}_6}$ is fitted, as the Fermi velocity of the surface bands, which determines the value of $\xi_\text{N}$, does not have a clear TI layer thickness dependence.\cite{SuYangXu2014NatPhys}

\begin{table}
\centering
\begin{tabular}{c |c|c|c}
\hline
\hline
\multirow{2}{*}{Characteristic lengths} & \multicolumn{3}{c}{SmB$_6$ layer thickness} \\
\cline{2-4}
 & 20 nm & 40 nm & 100 nm \\
\hlinewd{1.5pt}
$\xi_\text{N}$(2K) (nm)& $52\pm 1$ & 52$^*$ & 52$^*$ \\
\hline
$d_\text{N}$ (nm) & 20$^*$ & $8\pm 2$  & $ 10\pm 1$  \\
\hline
$\lambda_\text{N}(0,0)$ (nm) & $340\pm 2$ &$159\pm 2$ & $207\pm 2$  \\
\hline
\hline
\end{tabular}
\caption{\label{table2} Summary of the extracted characteristic lengths from the electrodynamic screening model for TI/SC bilayers for different SmB$_6$ layer thickness. All fits on the bilayers assume $\lambda_S(0) = 227$ nm which is obtained from the fitting on the single layer YB$_6$. Note that the values with the asterisk are fixed when the fitting is conducted. }
\end{table}

\par For the bilayers with $t_{\text{SmB}_6}=40$ and $100$ nm, $d_\text{N}$ is now set to be a free fitting parameter. As seen from Fig. \ref{fig:Fig3}(c) and (d), the resulting fit line gives $d_\text{N}=8\pm2$ nm, $\lambda_\text{N}(0,0)=159\pm2$ nm for the bilayer with $t_{\text{SmB}_6}=40$ nm, and $d_\text{N}=10\pm1$ nm, $\lambda_\text{N}(0,0)=207\pm2$ nm for the bilayer with $t_{\text{SmB}_6}=100$ nm. The estimated $d_\text{N} \approx 9$ nm is much smaller than $t_{\text{SmB}_6}$, which is consistent with the absence of induced order parameter in the top surface of 40 and 100 nm thick SmB$_6$ layers measured by point contact spectroscopy.\cite{SeunghunLee2019Nature} A summary of the estimated characteristic lengths $\xi_\text{N}$(2K), $d_\text{N}$, and $\lambda_\text{N}(0,0)$ for the case of 20, 40, and 100 nm thick SmB$_6$ layers on top of YB$_6$ is presented in Table. \ref{table2}.

\section{Discussion}
\par
\begin{figure}
		\includegraphics[width=1\columnwidth]{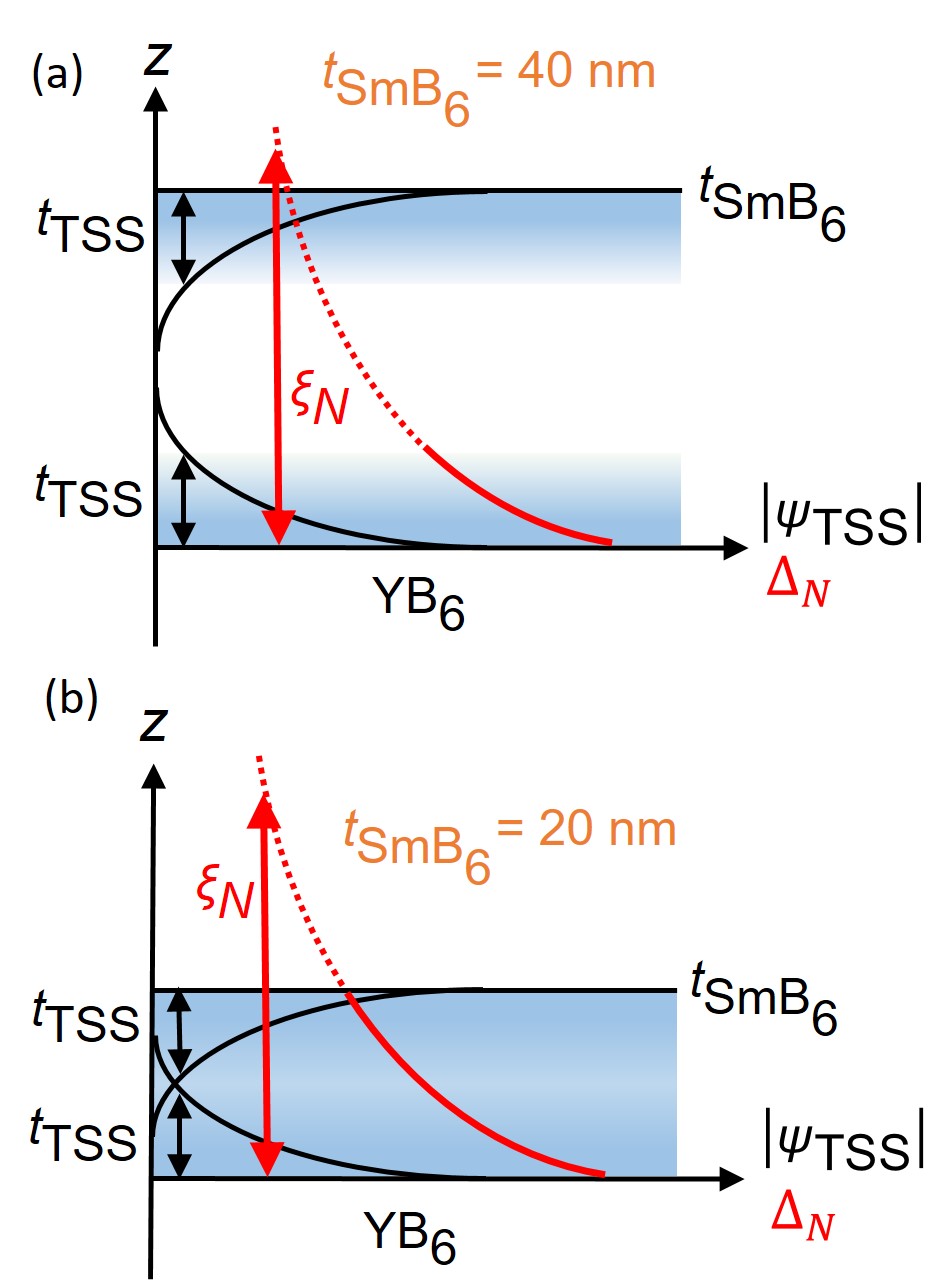}
		\caption{\label{fig:Fig4} Schematic view (not to scale) of the proposed position dependence of the surface states wavefunction $|\psi_\text{TSS}(z)|$ (black) and induced order parameter $\Delta_\text{N}(z)$ (red) in the SmB$_6$/YB$_6$ bilayer. The $|\psi_\text{TSS}(z)|$ is also visualized by the blue gradations. The sketches are based on the estimated proximity-coupled thickness $d_\text{N} \approx 9$ nm and the normal coherence length $\xi_\text{N}(2\text{K}) = 52$ nm for the case of $t_{\text{SmB}_6}$= (a) 40 nm, and (b) 20 nm. }
\end{figure}

 We now discuss the implications of these results and propose a microscopic picture for the proximity coupled bilayers. The important implication of the above results is the absence of Meissner screening in the bulk of proximity-coupled SmB$_6$, which is consistent with the existence of an insulating bulk region inside the SmB$_6$ layer. If the entire SmB$_6$ layer is conducting without an insulating bulk inside, the proximity-coupled thickness $d_\text{N}$ should be equal to $t_{\text{SmB}_6}$ for thicker films too, considering the long normal coherence length of $\approx52$ nm. In that case, as $t_{\text{SmB}_6}$ increases, one would expect a continuous evolution of stronger $\Delta\lambda(T)$ as seen in the Cu/Nb system (Fig. \ref{fig:Fig1}(c)), which is not observed in Fig. \ref{fig:Fig1}(b). Also, the estimated $d_\text{N}\approx9$ nm for the bilayers with $t_{\text{SmB}_6}$= 40 and 100 nm is much smaller than half of $t_{\text{SmB}_6}$. As illustrated in Fig. \ref{fig:Fig4}(a), this situation can only be explained if a thick insulating bulk region of $t_{\text{bulk}}\approx 22$ and $82$ nm exist in the bilayers with $t_{\text{SmB}_6}=$40 and 100 nm respectively.

\par This thick insulating bulk provides a spatial separation between the top and bottom surface conducting states, not allowing the order parameter to propagate to the top surface. Thus, only the bottom surface states are proximitized in the $t_{\text{SmB}_6}=$40 and 100 nm cases, and hence one can conclude that the proximitized thickness $d_\text{N}\approx9$ nm equals the thickness of the surface states $t_{\text{TSS}}$. Note that this confirmation of the presence of the insulating bulk in the TI layer cannot be made solely from the PCS study. Even if the PCS study observed the absence of the order parameter on the top surface of the TI layer (SmB$_6$ in this case), it could be either due to an insulating bulk, or due to a short normal coherence length $\xi_\text{N} <  t_{\text{SmB}_6}$. The large value of $\xi_\text{N}=52$ nm, which is larger than $t_{\text{SmB}_6}=40$ nm, rules out the latter scenario and confirms the presence of an insulating bulk inside the SmB$_6$ layers.

\par This picture is also consistent with the observation that the entire SmB$_6$ layer with $t_{\text{SmB}_6}=20$ nm is proximity-coupled (Fig. \ref{fig:Fig4}(b)); the top and the bottom conducting surface state wavefunctions are likely to be weakly overlapped based on $2t_{\text{TSS}}\approx t_{\text{SmB}_6}$ through the exponentially decaying profile (Fig. \ref{fig:Fig4}(b)). Thus the induced order parameter is able to reach to the top surface states, giving $d_\text{N}=20$ nm for this case. Although such overlap is expected to open a hybridization gap in the surface states, the fact that 20 nm SmB$_6$ on YB$_6$ is entirely proximity-coupled implies that the opened gap is much smaller than the energy difference between the Fermi level of SmB$_6$ and the Dirac point. Note that topological protection might not be affected by such weak hybridization, provided that the Fermi level is sufficiently far away from the Dirac point present in thick SmB$_6$.\cite{SuYangXu2014NatPhys}

\par Besides confirming the existence of an insulating bulk in the SmB$_6$ layer, the extracted fitting parameters based on the electromagnetic model provide an estimate for the important characteristic lengths such as $\xi_\text{N}$, $\lambda_\text{N}$, and $t_\text{TSS}$, as seen from Sec. \ref{Result}. These estimates can be utilized in designing a TI/SC device such as a vortex MBS device. $\xi_\text{N}$ determines the radius of the vortex core $r_v$. In the mixed state above the first critical field, $\lambda_\text{N}$ determines the maximum spacing $R_v$ between neighboring vortices in the vortex lattice.\cite{Tinkham1996} The ratio $r_v/R_v$ determines the overlap of the two adjacent MBSs. The overlap of the wavefunctions of the two MBSs results in intervortex tunneling, which splits the energy level of the MBSs away from the zero energy and make them trivial fermionic excitations,\cite{MCheng2010PRB}
\begin{equation}
\Delta E_\text{split} \sim \frac{1}{\sqrt{k_FR_v(\lambda_\text{N})}}\exp\left(-\frac{R_v(\lambda_\text{N})}{r_v(\xi_\text{N})}\right).
\end{equation} 
Therefore, information on $\xi_\text{N}$ and $\lambda_\text{N}$ helps to evaluate how secure the MBSs of a device will be against the intervortex tunneling.

\par $t_{\text{TSS}}$ determines a minimum required thickness of the device. If the thickness of the device is too thin ($t_{\text{SmB}_6}\sim t_{\text{TSS}}$), the wavefunction overlap between the top and bottom surface states becomes significant, which opens a large hybridization gap up to the Fermi level. As a result, the surface states lose not only the electric conduction but also lose the spin-momentum locking property.\cite{SuYangXu2014NatPhys} In this case, MBS is not hosted in the vortex core, and hence a thickness larger than the estimated $2t_{\text{TSS}}$ is recommended. These discussions show how the characteristic lengths extracted from the Meissner screening study serve as a guideline to design a vortex MBS device with TI/SC bilayer systems.

\section{Conclusion}
\par In summary, a microwave Meissner screening study is introduced and utilized to investigate the spatially dependent electrodynamic screening response and the corresponding properties of the TI/SC bilayers. The advantages of the study in investigating the properties of a TI/SC system is demonstrated by the measurement and modeling of the temperature dependence of the screening with systematic TI-layer thickness variation. The study goes beyond the surface response to examine the screening properties of the entire TI layer, and uncovers the existence of an insulating bulk in the TI layer conclusively. Also, the study provides an estimate for characteristic lengths of the TI/SC bilayer, which sheds light on the design of a vortex MBS device providing guidelines for the radius of the vortex core, the energy level splitting due to the intervortex tunneling, and the thickness of the device. With its versatile capabilities, the microwave Meissner screening study can serve as a standard characterization method for a variety of TI/SC systems before using them as building blocks in topological quantum computation.

\begin{acknowledgments}
The authors thank Yun-Suk Eo and Valentin Stanev for helpful discussions. This work is supported by NSF grant No. DMR-1410712, DOE Office of High Energy Physics under Award No. DE-SC 0012036T (support of S.B.), Office of Basic Energy Science, Division of Material Science and Engineering under Award No. DE-SC 0018788 (measurements), ONR grant No. N00014-13-1-0635, AFOSR grant No. FA 9550-14-10332 (support of S.L., X.Z., and I.T.), and the Maryland Center for Nanophysics and Advanced Materials.
\end{acknowledgments}

\appendix
\section{Sample properties}
\subsection{Thin film bilayer preparation} \label{SampleGrowth}
\par SmB$_6$/YB$_6$ bilayers were prepared by an in-situ sequential sputtering process (i.e., without breaking vacuum) to secure the ideal superconducting proximity effect which is a prerequisite for the current study and analyses.\cite{SeunghunLee2016PRX} SmB$_6$ and YB$_6$ share the same crystal structure with almost the same lattice constant ($\approx$ 4.1 \AA), which allows the fabrication of bilayers by sequential high-temperature growth under the same conditions. YB$_6$ is a superconducting rare-earth hexaboride and it has been reported that slight boron deficiency improves the superconducting transition temperature ($T_c$) of YB$_6$.\cite{SeunghunLee2019Nature} Thus, for this study, slightly boron deficient YB$_6$ films (B/Y = 5.6) were used as the superconducting layers. YB$_6$ thin films were deposited on Si(001) substrates. To remove the native oxide layer on the Si substrate, we treated it with hydrofluoric acid (HF) before the thin film deposition. The base pressure of the deposition system was $2\times 10^{-8}$ Torr. The deposition process was performed at 860 $^{\circ}\mathrm{C}$ under a pressure of 10 mTorr adjusted by Ar gas (99.999 \%). The thickness of YB$_6$ layers was fixed to be 100 nm. The subsequent SmB$_6$ deposition was performed under the same temperature and pressure conditions, and an additional sputtering of B target was employed to compensate B deficiency which is present in the films fabricated by the sputtering of a stoichiometric SmB$_6$ target.\cite{JieYong2014APL,SeunghunLee2016PRX} The compositions (i.e., stoichiometry) of YB$_6$ and SmB$_6$ thin films were examined with wavelength dispersive spectroscopy (WDS) measurements. The thicknesses of bilayers were confirmed with cross-sectional scanning electron microscopy (SEM) measurements.

\subsection{Validity of the estimated magnetic penetration depth of the YB$_6$ thin film } \label{YB6 penetration depth}
In the main text (Fig. 2(a)),  the model fit gives $\lambda_S(0)=227 \pm 2$ nm (and $2\Delta(0)/k_BT_c=3.66 \pm 0.01$) for the YB$_6$ thin film with thickness of 100 nm.
This estimate is larger than the value $\lambda_S(0)\approx 134$ nm measured by muon spin rotation study from a single crystal YB$_6$ sample\cite{Kadono2007PRB} with higher $T_c=6.94$ K (and $2\Delta(0)/k_BT_c=3.67$). This is reasonable considering that the higher $T_c$ implies a longer mean free path $l_{mfp}$\cite{Sluchanko2017PRB}, and shorter $\lambda_S(0)$ through the relation $\lambda_S(0)=\lambda_L(0)\sqrt{1+\xi_0/l_{mfp}}$\cite{Tinkham1996} where $\lambda_L(0)$ is London penetration depth at $T=0$ K and $\xi_0$ is BCS coherence length of the superconductor.

\subsection{Validity of the extracted characteristic lengths of the SmB$_6$/YB$_6$ bilayers}
\par To confirm the validity of the estimated values of the characteristic lengths of the SmB$_6$/YB$_6$ bilayers obtained in Sec. \ref{Result}, one of the parameters $\xi_\text{N}$ is converted to the Fermi velocity $v_F$, whose value has been reported from other measurements on SmB$_6$. From the clean limit relation $\xi_\text{N} = \hbar v_F/2\pi k_B T$, one arrives at $v_F = 8.5\times 10^4$ m/s. As seen from Table. \ref{table1}, this value is similar to the values obtained from the ARPES and DC transport measurements. However, the $v_F$ values from theory and STM are an order of magnitude smaller. Recent DFT calculation accompanied by STM measurements\cite{Pirie2018ArXiv,Matt2018ArXiv} and an independent theoretical calculation\cite{Alexandrov2015PRL} show that the discrepancy can be explained by termination-dependent band bending at the surface of SmB$_6$. The value of $\xi_\text{N} (2\text{K})$ is also directly compared to that obtained from the DC transport study on Nb/SmB$_6$ bilayers.\cite{SeunghunLee2016PRX} The transport study has estimated a smaller value (9 nm) compared to our result (52 nm). This could be due to the differences in the grain size.

\begin{table}
\centering
\begin{tabular}{c @{\qquad} l @{\qquad}l}
\hline
\hline
 & This work & previous work \\
\hline
$\xi_\text{N}$(2K) (nm)& $52$ & $9$\cite{SeunghunLee2016PRX} \\
 & (clean limit) & (dirty limit) \\
\hline
$v_F$ & $8.5$ & 4\cite{Neupane2013NatComm,Jiang2013NatComm} (ARPES)  \\
($10^4$ m/s)&  & 9\cite{SeunghunLee2016PRX} (transport) \\
& & 0.6\cite{Pirie2018ArXiv} (STM) \\
& & 0.4\cite{Roy2014PRB} (theory) \\
\hline
$t_{\text{TSS}}$ (nm)& $\approx 9$ & 6\cite{SeunghunLee2016PRX} (transport)  \\
 & & 32\cite{LiuTao2018PRL} (spin pumping) \\
\hline
\hline
\end{tabular}
\caption{\label{table1} Characteristic lengths ($\xi_\text{N}$ and  $t_\text{TSS}$) of the SmB$_6$/YB$_6$ bilayers and derived property ($v_F$) for SmB$_6$ obtained from the microwave Meissner screening study compared to those from previous studies.}
\end{table}

\section{Dielectric resonator setup} \label{AppDielectricResonator}
\par
\begin{figure}
		\includegraphics[width=1\columnwidth]{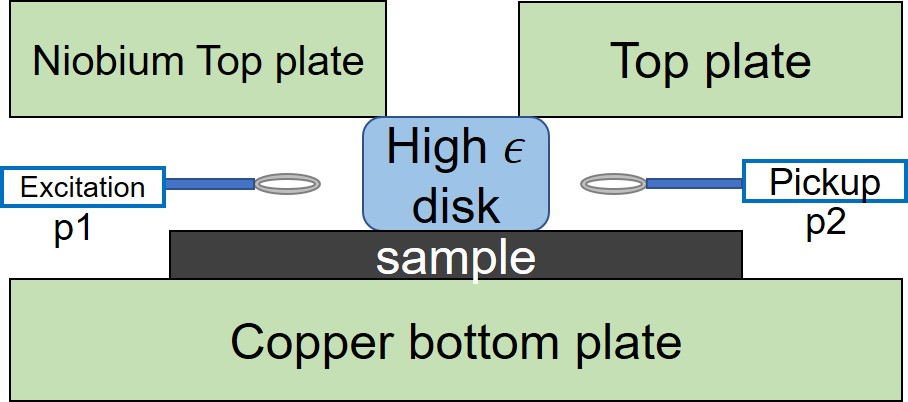}
		\caption{\label{fig:FigSM1} Schematic cross-section diagram of the dielectric resonator setup for a microwave transmission resonance with a sample. }
\end{figure}

 The dielectric resonator setup was originally developed to study dielectric properties of materials\cite{HakkiColeman1960} and subsequently used to characterize microwave properties of high-$T_c$ cuprate films.\cite{Klein1992JSuper,Mazierska1998IEEE,Pompeo2007JSuper} The comprehensive details  of the dielectric resonator used in this work can be found in Ref. \cite{SeokjinBae2019RSI} Here, a summary of the key features is introduced for the reader's convenience. The resonator consists of a top and bottom metallic plate which confine the microwave field inside the resonator just as in a cavity (Fig. \ref{fig:FigSM1}). A disk with high dielectric constant, which is placed on top of a superconducting thin film sample, concentrates the incident microwave fields injected from the excitation loop (p1 of Fig. \ref{fig:FigSM1}) in the disk and generates a microwave resonance at certain frequencies $f_0$. These resonant frequencies $f_0$ are determined mainly by the dimension and the dielectric constant of the disk. In our setup, a 3 mm diameter, 2 mm height rutile (TiO$_2$) disk is used as the dielectric disk. Rutile is chosen as the dielectric material for the resonator because it has very high dielectric constant ($\epsilon_{c} > 250$, $\epsilon_{a,b} > 120$ where a, b are the in-plane crystallographic axes and c is the out-of-plane axis) compared to those of sapphire ($\epsilon_{a,b,c}\sim10$) or other dielectric materials. The high dielectric constant of the rutile helps to minimize the size of the disk, while maintaining the resonant frequencies in the microwave regime. The smaller the measurement area is, the more likely the sample will have homogeneous properties. Among the resonant modes generated by the dielectric resonator, the TE$_{011}$ mode ($\mathtt{\sim}11$ GHz) induces a radial magnetic field and a circulating screening current on the sample surface. This circulating current helps to support the microwave transmission resonance. If there occurs any change of the sample properties such as superfluid density, that change can be studied through the change of the microwave transmission resonance. Note that the typical value of the quality factor of the TE$_{011}$ mode in this work is on the order of $10^4$. The simulated (HFSS) microwave magnetic field at the surface of the sample for the TE$_{011}$ mode is $\approx 8 \mu T$ when the input microwave power $P_{in}$ is $-20$ dBm. In this range of $P_{in}$, the resonance frequency does not show $P_{in}$ dependence, showing that the sample is in the linear response regime in terms of the microwave magnetic field.

\section{Measurement of the effective penetration depth}
\subsection{Determining resonance frequency and corresponding effective penetration depth} \label{lambda}
\par Microwave transmission data $S_{21}(f)$ near the resonance is fitted with the phase versus frequency fitting procedure,\cite{Petersan1998JAP} to precisely determine the resonance frequency $f_0$. Measurement and fitting of $S_{21}(f)$ data are repeated for different temperatures. From this, the temperature dependence $\Delta f_0(T) = f_0(T)-f_0(T_{ref})$ can be acquired. This temperature dependence of the resonance frequency can be converted to that of the effective penetration depth of a superconducting thin film sample by\cite{Hein1999,BBJin2002PRB,Ormeno2002PRL}
\begin{equation}
\Delta\lambda_{eff}(T) = -\frac{G_{geo}}{\pi \mu_0}\frac{\Delta f_0(T)}{f_0^2(T)}.
\end{equation}
Here, $G_{geo}$ = $\omega\mu_0\int_V dV|H(x,y,z)|^2$ / $\int_S dS |H(x,y)|^2$ = $225.3$ $\Omega$ is the geometric factor calculated numerically using the field solution inside the resonator for TE$_{011}$ mode derived by Hakki et al\cite{HakkiColeman1960}.

\subsection{Determining error bars for the effective penetration depth and estimated fit parameters} \label{ErrorbarDet}
\par The error bar in the effective penetration depth $\Delta\lambda_{eff}(T)$ is determined by the error bar of determination of the resonance frequency $f_0(T)$. The error bar of the $f_0$ is determined by a deviation of $f_0$ from the estimated value, which increases the root-mean-square error $\sigma$ of the fit by 5\%. The main source of the error bar of $f_0$ is the noise in $S_{21}(f)$ data. If the signal to noise ratio of $S_{21}$ is large (small) which makes the $S_{21}(f)$ curve well- (poorly-) defined, $f_0$ can have a narrower (wider) range of values while giving fits with similar values of $\sigma$. Once the error bar of $f_0$ is determined, with the standard error propagation from the relation between $\Delta\lambda_{eff}(T)$ and $f_0(T)$, the error bar in the $\Delta\lambda_{eff}(T)$ data is estimated. The error bar for the estimated fit parameters ($\xi_N(T_0)$, $\lambda_N(0,0)$, and $d_N$) obtained from fitting $\Delta\lambda_{eff}(T)$ data are determined by a deviation from the estimated value which increases $\sigma$ by 5\%.

\section{Further remarks on the electromagnetic screening model} \label{Screening model equation}
\subsection{Boundary conditions}
\par Although explained in detail in Ref.\cite{Pambianchi1994PRB}, for the reader's convenience, the equation and the boundary conditions for the magnetic field inside a proximity-coupled bilayer are described below. First, by combining Maxwell's equations with London's equation, one can obtain an equation for the tangential magnetic field for the bilayer
\begin{equation}\label{fieldEq}
\frac{d^2H(z)}{dz^2}+\frac{2}{\lambda_{N,S}(z)}\frac{d\lambda_{N,S}(z)}{dz}\frac{dH(z)}{dz}-\frac{1}{\lambda_{N,S}^2(z)}H(z)=0.
\end{equation}
The boundary conditions for the tangential magnetic field for the geometry shown in Fig. 1 of the main article are as follows,
\begin{gather}
H(d_N)=H_0 \textrm{,  (top surface)}\\
H(-d_S)=0 \textrm{,  (bottom surface)}\\
H(0^+)=H(0^-) \textrm{,  (interface)} \\
\lambda_N^2(0,T)\frac{dH(z)}{dz}|_{z=0^+}=\lambda_S^2(0,T)\frac{dH(z)}{dz}|_{z=0^-},
\end{gather}
where $d_N\leq t_{SmB_6}$ is the proximity-coupled thickness of the normal layer and $d_S = t_{YB_6}$ is the thickness of the parent superconductor. The last boundary condition is a continuity condition for the superfluid velocity at the interface.

\subsection{Field solutions}
\par With Eq.(\ref{fieldEq}) and the approximated spatial profile of the induced order parameter in the normal layer $\Delta_N(z,T) = \Delta_N(0,T)e^{-z/\xi(T)}$ and the normal penetration depth $\lambda_N(z,T)=\lambda_N(0,T)e^{+z/\xi_N(T)}$, one can obtain the spatial profile of the magnetic field in the normal and superconducting layer as follows:\cite{Pambianchi1994PRB}
\begin{gather}
H_N(z,T) = ApI_1(p) + BpK_1(p) \textrm{,  ( $0\leq z\leq d_N$)} \\
H_S(z,T) = Ce^{z/\lambda_S} + De^{-z/\lambda_S} \textrm{,  ($-d_S\leq z\leq 0$)},
\end{gather}
Here, the parameter $p$ is defined as $p(z,T) = (\xi_N(T)/\lambda_N(z,T)) e^{-z/\xi_N(T)}$ and $I_1$, $K_1$ are the modified Bessel functions of the first, second kind. The coefficients $A,B,C,D$ can be calculated using the boundary conditions. The corresponding spatial profile of the current density can be obtained from $z$ derivative of the magnetic field profile. After all the coefficients are obtained, the spatial profiles of the magnetic field and the current density of a normal/superconductor bilayer are fully determined. When calculating the inductance, the microwave loss is ignored so that the supercurrent density of the bilayer is approximated as the total current density $J_s\simeq J$. This is a valid approximation since the temperature range of the measurement (0$\mathtt{\sim}1.6$ K) is well below $T_c$ of the bilayer ($\mathtt{\sim}5.86$ K) and the microwave photon energy ($\mathtt{\sim}0.044$ meV) is much lower than the zero temperature superconducting gap of the YB$_6$ ($>1$ meV).\cite{Kadono2007PRB} 

\par 

\bibliography{Ref_Meissner_TIS}

\end{document}